\documentclass{PoS}
\usepackage{epstopdf}
\title{ALICE status and plans}

\ShortTitle{ALICE status and plans}

\author{Evgeny KRYSHEN for the ALICE Collaboration
         \\
        Petersburg Nuclear Physics Institute (PNPI), \\
 National Research Center ''Kurchatov Institute", Gatchina, 188300, Russia\\
        E-mail: \email{evgeny.kryshen@cern.ch}}

\def\jpsi{J$/\psi$ }

\abstract{The ALICE experiment has been running successfully since 2010 and made an impressive progress towards understanding of hot and dense QCD matter produced in heavy ion collisions at LHC energies. Recent results on identified particle spectra, azimuthal anisotropy, heavy flavour and quarkonium production in Pb--Pb collisions at $\sqrt{s_{NN}} = 2.76$ TeV are presented. First results on p--Pb collisions and ALICE upgrade plans are briefly reviewed.
}

\FullConference{LHC on the March\\
		 20-22 November 2012\\
		 Institute for High Energy Physics, Protvino, Moscow region, Russia }

\begin{document}
\section{Introduction}
ALICE is the largest dedicated heavy ion experiment at LHC aimed to study hot and dense QCD matter produced in heavy ion collisions. These studies are important to advance our knowledge on the early Universe evolution and poorly understood QCD subjects such as confinement and chiral symmetry restoration. The matter produced in heavy ion collisions was initially thought to behave as a weakly interacting gas of deconfined quarks and gluons due to onset of the asymptotic freedom regime of QCD. Surprisingly, the medium produced at RHIC and LHC energies manifested itself as an almost perfect liquid with a short mean free path, high collectivity effects and large parton energy loss. The main ALICE goal now is to obtain a detailed quantitative description of the properties and dynamics of the produced matter at LHC. 
\begin{figure}[b]
\includegraphics[width=13cm]{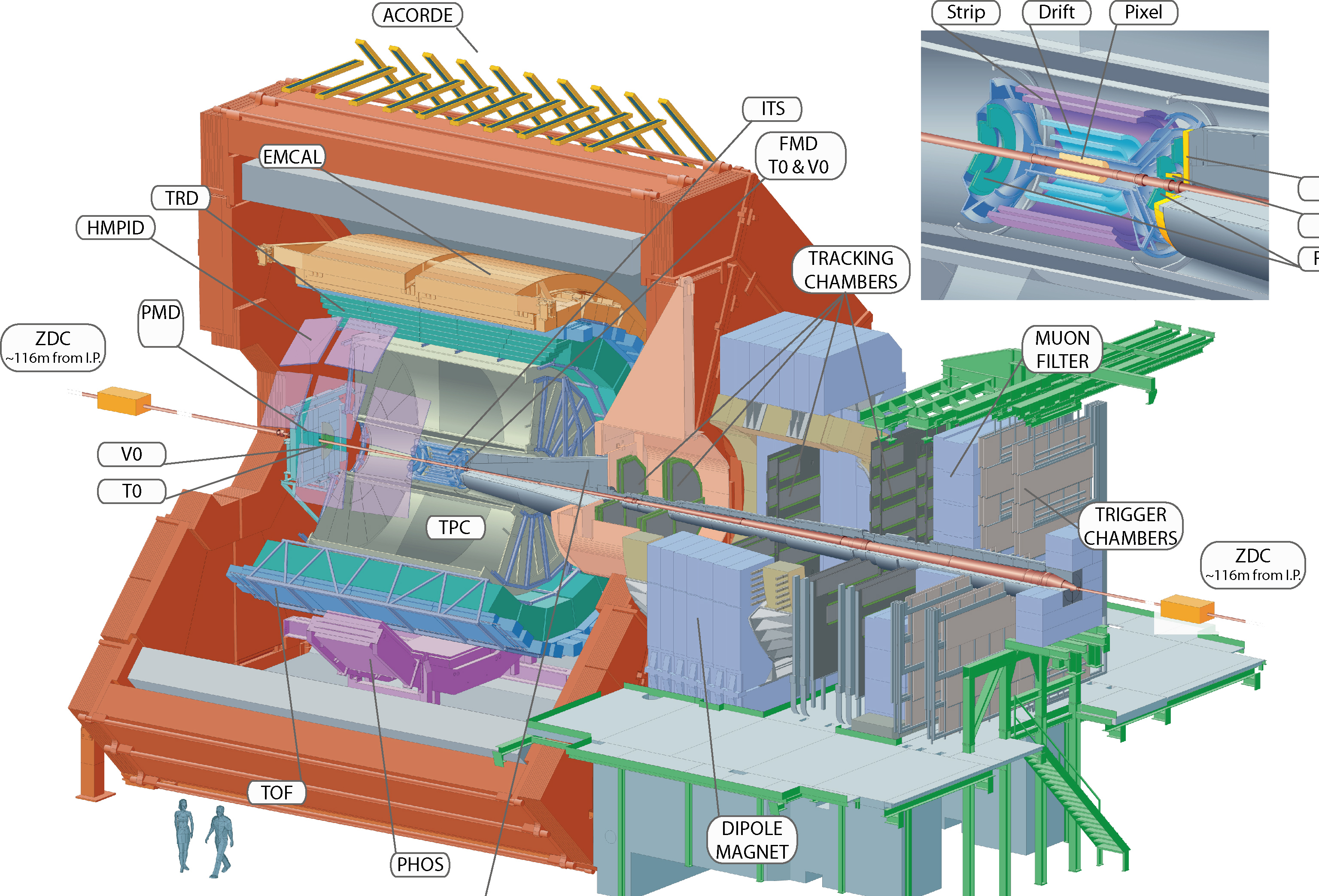}
\caption{Schematic layout of the ALICE detector.}
\label{fig:setup}
\end{figure}

ALICE experimental setup, shown in fig.~\ref{fig:setup}, consists of two major parts: a central barrel and a forward muon spectrometer. The central barrel is embedded in a solenoid magnet with a magnetic field of 0.5 T and covers a pseudorapidity range $|\eta| < 0.9$. It consists of the Inner Tracking System (ITS) with 6 layers of high resolution silicon detectors, a cylindrical Time Projection Chamber (TPC) and three particle identification detectors: the Transition Radiation Detector (TRD), the Time-Of-Flight detector (TOF) and the Cherenkov detector (HMPID). Two calorimeters, a high resolution photon spectrometer (PHOS) and an electro-magnetic calorimeter (EMCAL), are located at midrapidity and cover only a fraction of azimuthal acceptance. The muon spectrometer consists of a set of absorbers, a dipole magnet, five tracking and two trigger stations allowing to detect muons in the range $-4 < \eta -2.5$. Several detectors at forward rapidities  (T0, V0, ZDC, FMD, PMD) serve for the triggering and multiplicity measurements. Further details on the ALICE experimental setup can be found in~\cite{alice}.

During the first heavy ion runs in 2010 and 2011, ALICE recorded Pb--Pb collisions at $\sqrt{s_{NN}} = 2.76$ TeV corresponding to integrated luminosity of about 10 $\mu$b$^{-1}$ and 100 $\mu$b$^{-1}$ respectively. Recent results based on this data are briefly described in the following.

\section{Identified particle spectra}

ALICE provides unique tracking, vertexing and particle identification capabilities down to very low momenta of about 100 MeV/$c$ where particle spectra are mainly governed by the radial expansion of the medium. ALICE measurements of the low-momentum proton, pion and kaon $p_{T}$ spectra in central Pb--Pb collisions agree with hydrodynamic predictions within 20\% supporting hydrodynamic interpretation of the data at LHC~\cite{spectra}. Hydro-inspired blast wave fits to these spectra allowed to extract mean collective velocity of the transverse expansion which was found to be about 65\% of speed of light, 10\% higher than at RHIC and in good agreement with the observed tendency from the RHIC energy scan. 

\begin{figure}[b]
\includegraphics[width=7cm]{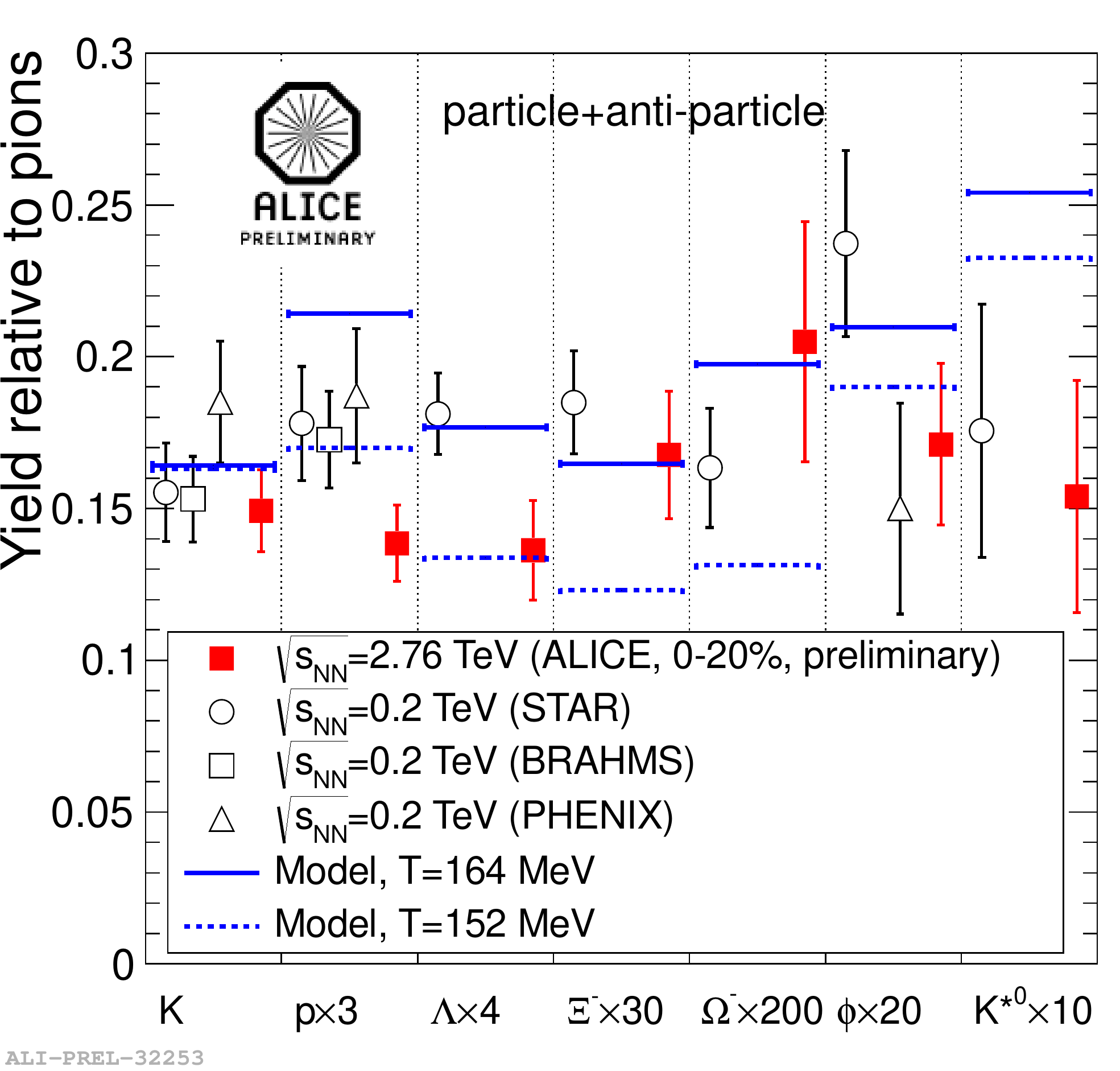}
\includegraphics[width=8.1cm]{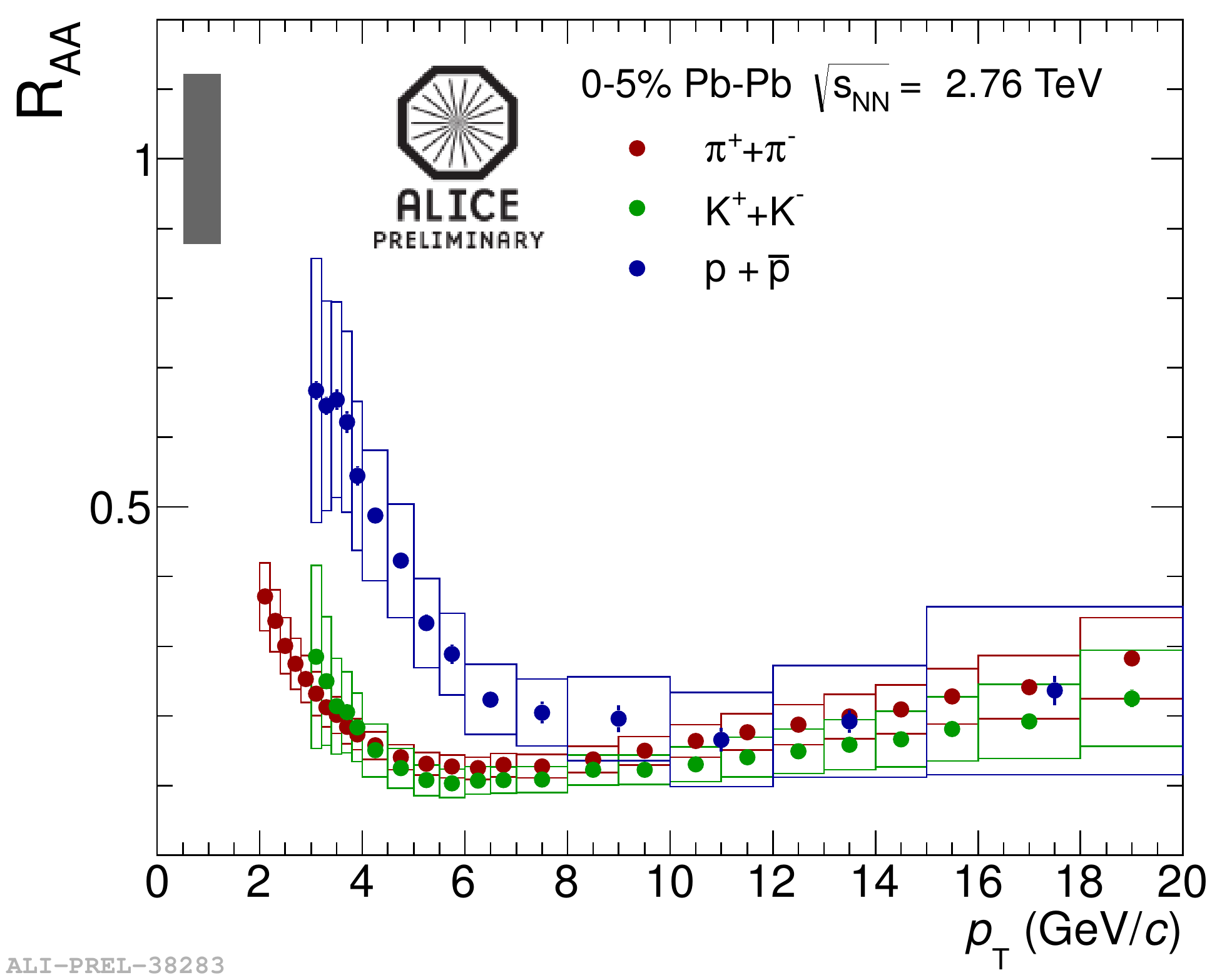}
\caption{Particle ratios compared to RHIC results and thermal model predictions (left) and nuclear modification factors for pions, protons and kaons (right) in central Pb--Pb collisions at $\sqrt{s_{NN}} = 2.76$ TeV.}
\label{termal}
\end{figure}

Integrated particle yields are shown in fig.~\ref{termal} (left) relative to pion yields. At AGS, SPS and RHIC, relative particle abundances were successfully described in terms of the thermal model with only two parameters: the chemical freeze-out temperature $T_{\rm ch}$ and the baryochemical potential $\mu_{B}$~\cite{andronic}.  However, thermal fits to ALICE data are rather poor ($\chi^2 = 39.6$ per 9 d.o.f.) and yield unexpectedly low $T_{\rm ch} = 152 \pm 3$ MeV mainly because $p/\pi$ and $\Lambda/\pi$ ratios appeared to be much lower than at RHIC. Excluding protons from the fit would significantly improve the agreement with ALICE data and yield $T_{\rm ch} = 164$ MeV close to the value extrapolated from lower energies. Arguably, the significant deviation from the usual thermal ratios can be explained by the final-state interactions in the hadronic phase, in particular via antibaryon-baryon annihilation~\cite{Becattini2012,Stenheimer2013}.

The nuclear modification factors ($p_{\rm T}$ spectra in Pb--Pb normalized to pp spectra and properly scaled by the number of binary nucleon--nucleon collisions)  for identified hadrons have been also measured by ALICE and are shown in fig.~\ref{termal} (right) for central Pb--Pb collisions. High $p_{\rm T}$ region (above 10 GeV/$c$) is dominated by particles from jet fragmentation. According to fig.~\ref{termal} (right) the particle production  at high $p_{\rm T}$ is suppressed by factor 5 for all particle species meaning that the jet hadrochemistry is not modified significantly with respect to pp collisions. In contrast, in the intermediate $p_{\rm T}$ range heavier particles exhibit weaker suppression which might be a consequence of the strong radial flow.

\section{Anisotropic flow}
Initial spatial asymmetry in non-central heavy ion collisions results in anizotropic momentum distributions (flow) of the produced particles providing unique tools to investigate collision dynamics. The flow is usually studied in terms of Fourier coefficients of azimuthal particle distributions and is found to be in good agreement with hadrodynamical calculations with very low viscosity both at RHIC and LHC energies~\cite{flow}. 

\begin{figure}[b]
\includegraphics[width=7.5cm]{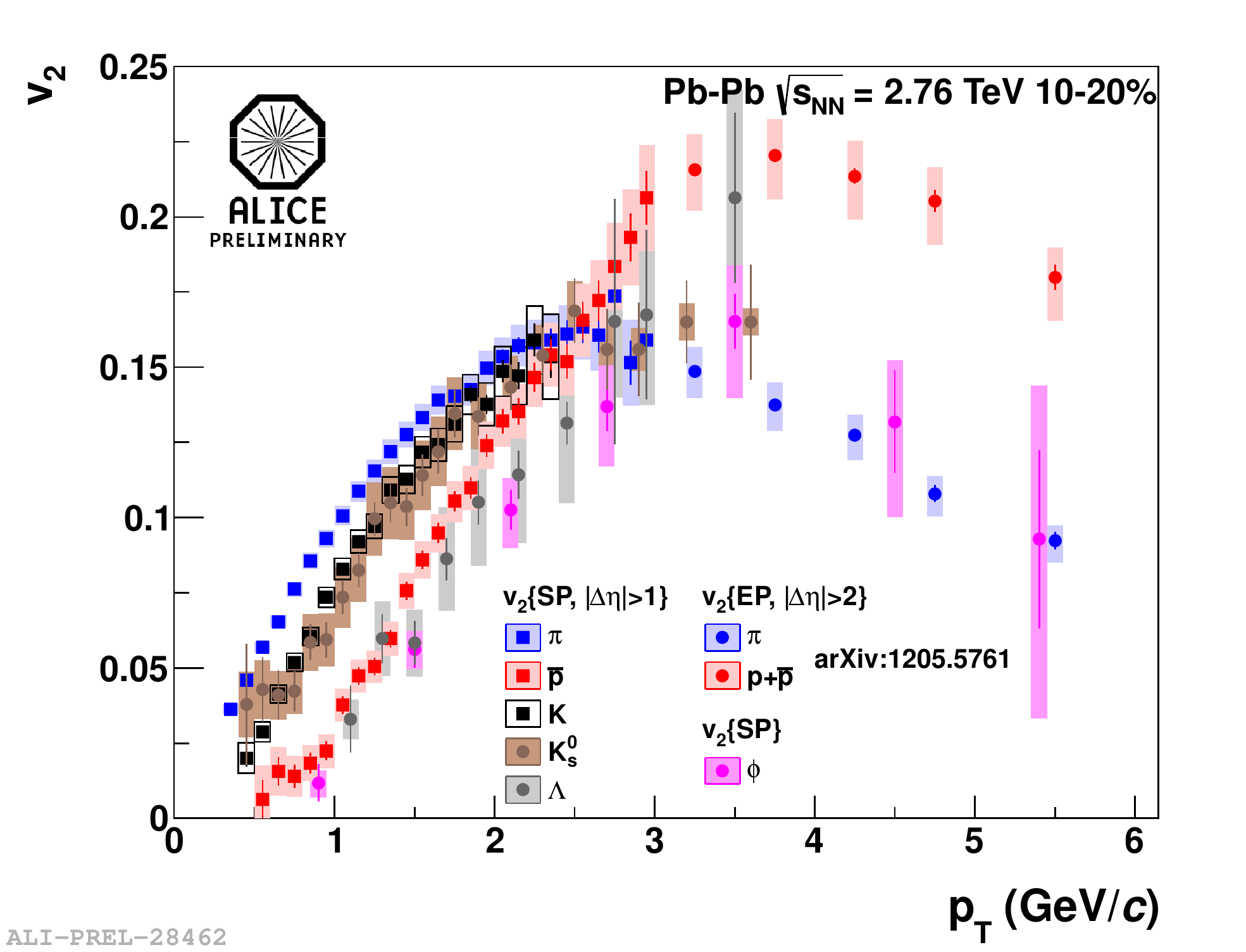}
\includegraphics[width=7.5cm]{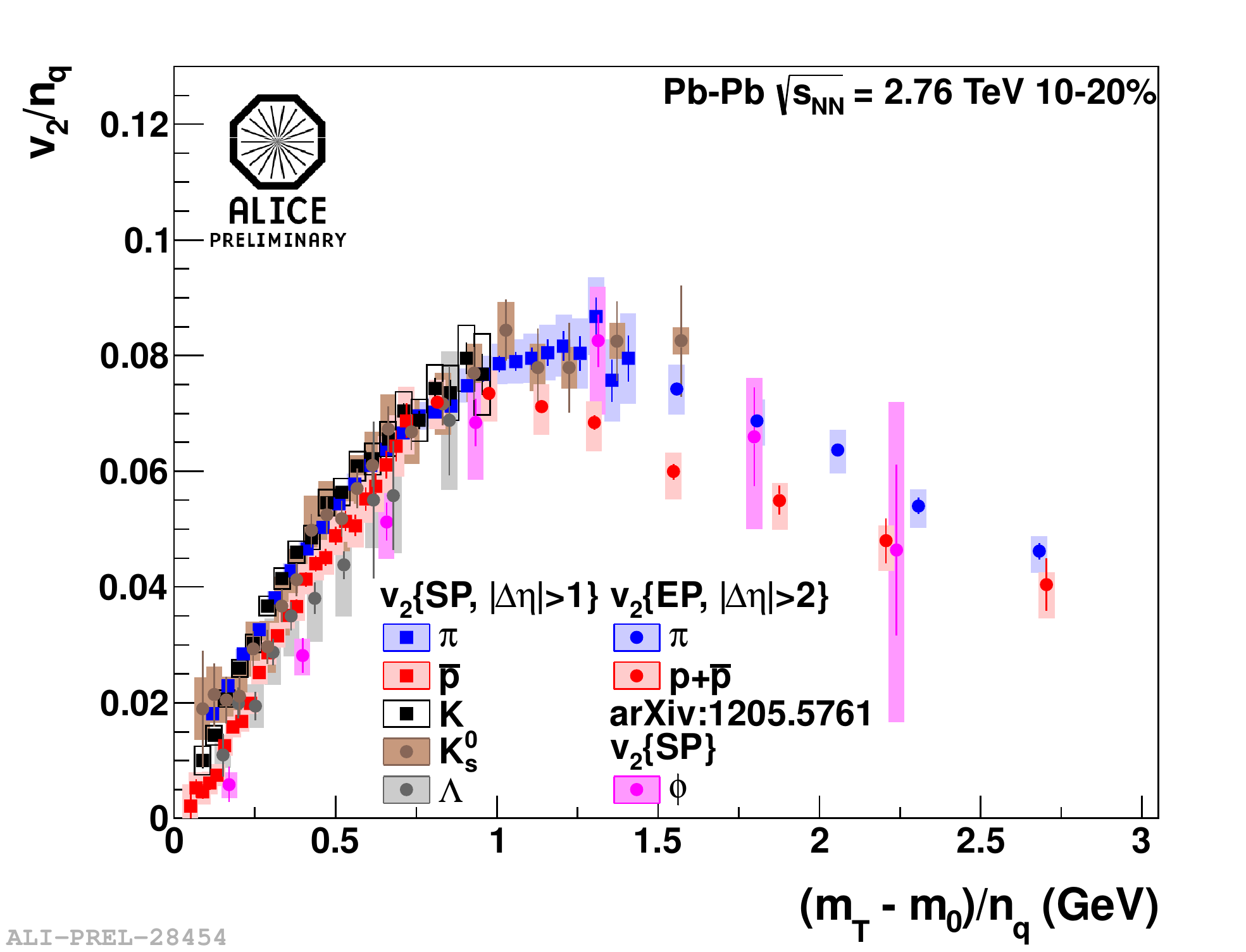}
\caption{$v_2$ as function of transverse momentum (left) and $v_2$ per constituent quark as a function of transverse kinetic energy  per constituent quark (right) for different particle species in 10--20\% centrality class.}
\label{fig:flow}
\end{figure}

Anisotropic flow of identified particles also provides valuable details on the dynamics and properties of the produced medium. Fig.~\ref{fig:flow} (left) represents ALICE results on the elliptic-flow coefficient $v_2$ for identified particles as function of $p_{\rm T}$ suggesting an existence of several regions in
the transverse momentum space with distinctly different underlying physics. At low $p_{\rm T}$, clear mass ordering is observed in qualitative agreement with hydrodynamic models in which heavier particles are pushed towards higher $p_{\rm T}$. Additional hadronic rescattering phase provides even better agreement~\cite{heinz}. At $p_{\rm T} > 3$ GeV/$c$, this tendency starts to change and $v_2$ for identified particles splits into meson and baryon bands. Such a splitting, also seen at RHIC at high $p_{\rm T}$, is naturally explained in the quark coalescence model, the so-called Number of Constituent Quark (NCQ) scaling~\cite{voloshin}. Observation of such a scaling is very important since hadronization via coalescence means that the system evolved from the deconfined phase. A better check of the NCQ scaling is presented in fig.~\ref{fig:flow} (right) where $v_2$ normalized to the number of constituent quarks $n_q$ is shown as a function of the transverse kinetic energy $m_{\rm T} - m$ ($m^2_{\rm T} = m^2+ p_{\rm T}^2$) per constituent quark. The NCQ scaling is not as good as at RHIC: there is a residual splitting at low $p_{\rm T}$ and differences in $v_2/n_{q}$ up to 20\% at higher $p_{\rm T}$ where particle mass is negligible.

\section{Heavy flavour}
\begin{figure}[h]
\includegraphics[height=6.5cm]{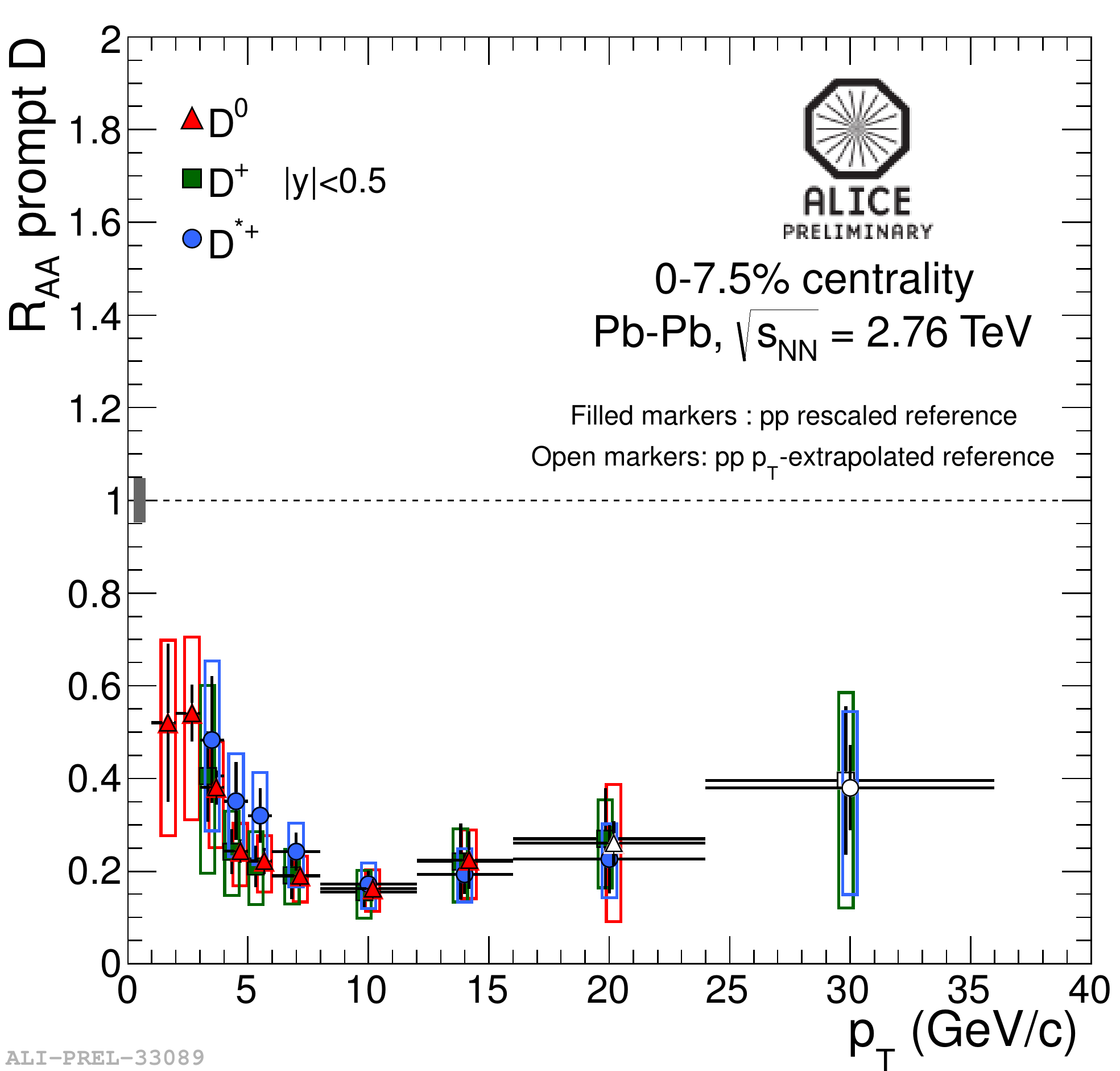}
\includegraphics[height=6.9cm]{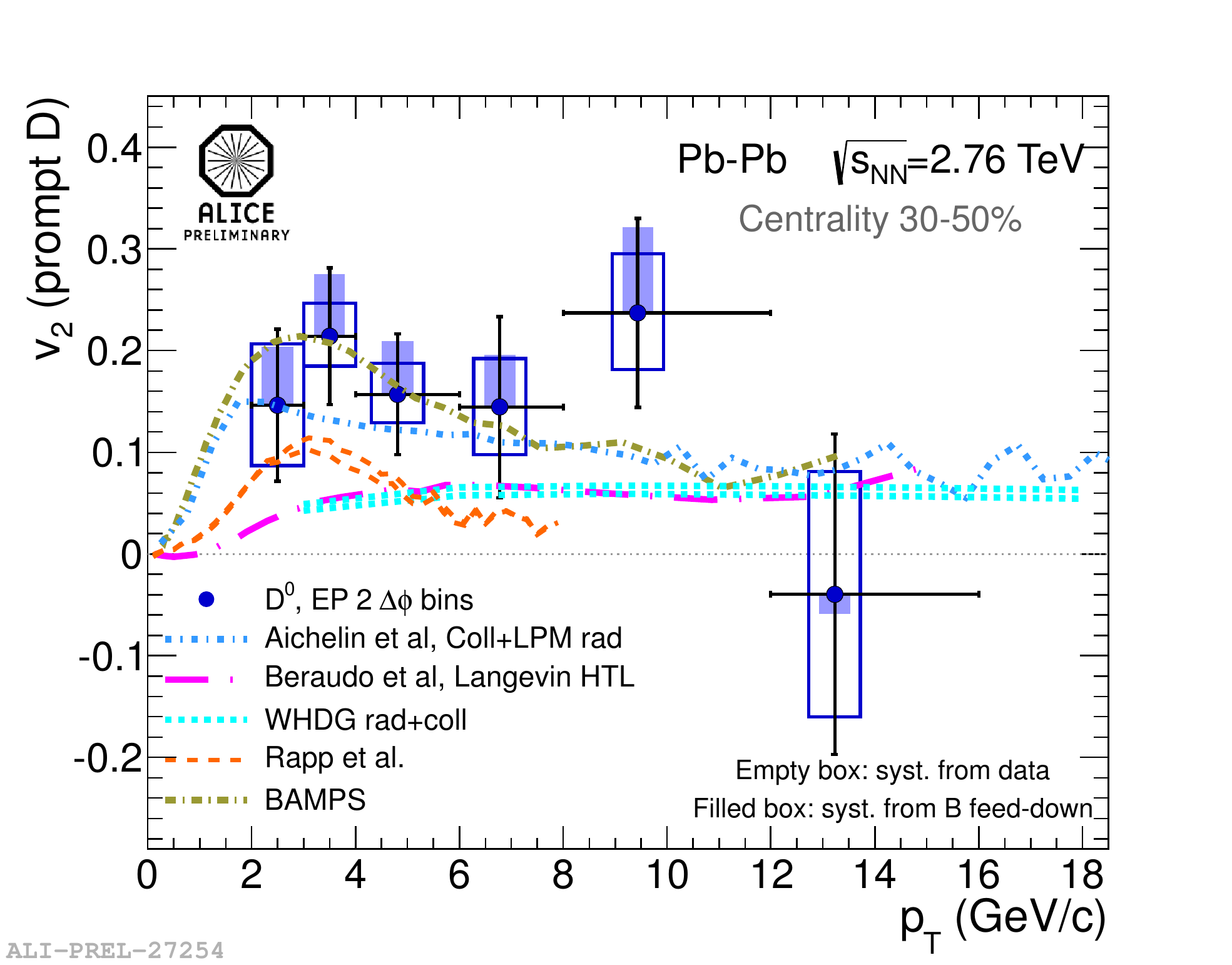}
\caption{Nuclear modification factors for $D^0$, $D^+$ and $D^{*+}$ mesons in central Pb--Pb collisions (left) and $v_2$ for prompt $D^0$ mesons in 30--50\% centrality bin (right) as functions of $p_{\rm T}$.}
\label{fig:d}
\end{figure}

Charm quarks are abundantly produced at LHC providing a powerful tool to study both the properties of the medium and the transport properties of charm quarks. In-medium energy loss for heavy quarks is expected to be smaller than for light quarks and gluons due to color charge and gluon-bremsstrahlung dead cone effects~\cite{armesto,dokshitzer}. Fig.~\ref{fig:d} (left) presents ALICE results on the nuclear modification factors for $D^0$, $D^+$ and $D^{*+}$ mesons. The observed suppression is similar for different meson species and reaches factor 5 at $p_{\rm T} \sim 10$ GeV/$c$ being very close to the average charged particle suppression and providing an indication of no strong color charge dependence of the in-medium energy loss. Similar suppression pattern have been also observed in semi-leptonic decay channels of heavy flavours.

The observed elliptic flow of prompt $D^{0}$ mesons, shown in fig.~\ref{fig:d} (right), is also comparable with $v_2$ of light hadrons if favour of possible thermalization of charm quarks in the hot QCD medium created in heavy-ion collisions at LHC. 

\section{Charmonium}
30 years ago charmonium suppression was proposed as the best signature of deconfined phase since $c\bar{c}$ states have to melt at high temperature due to color screening effects in QGP~\cite{matsui}. Large suppression was indeed observed at SPS and RHIC, however, with high multiplicity of heavy quarks at LHC one has to consider another scenario: enhancement of bound $c\bar{c}$ states via regeneration in thermalized QGP medium~\cite{thews} or during hadronization~\cite{pbm}. 

\begin{figure}[hb]
\includegraphics[width=7.5cm]{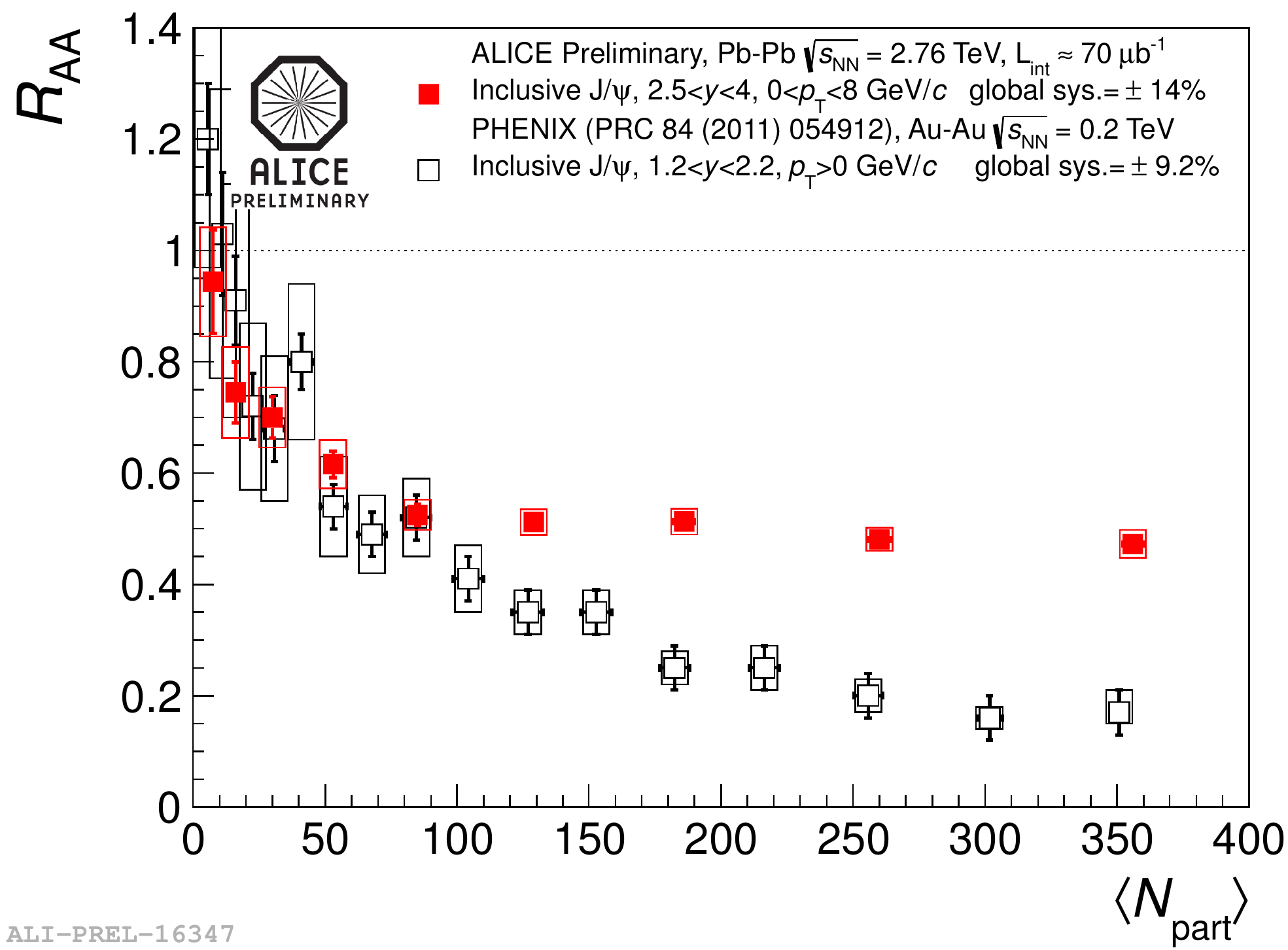}
\includegraphics[width=7.5cm]{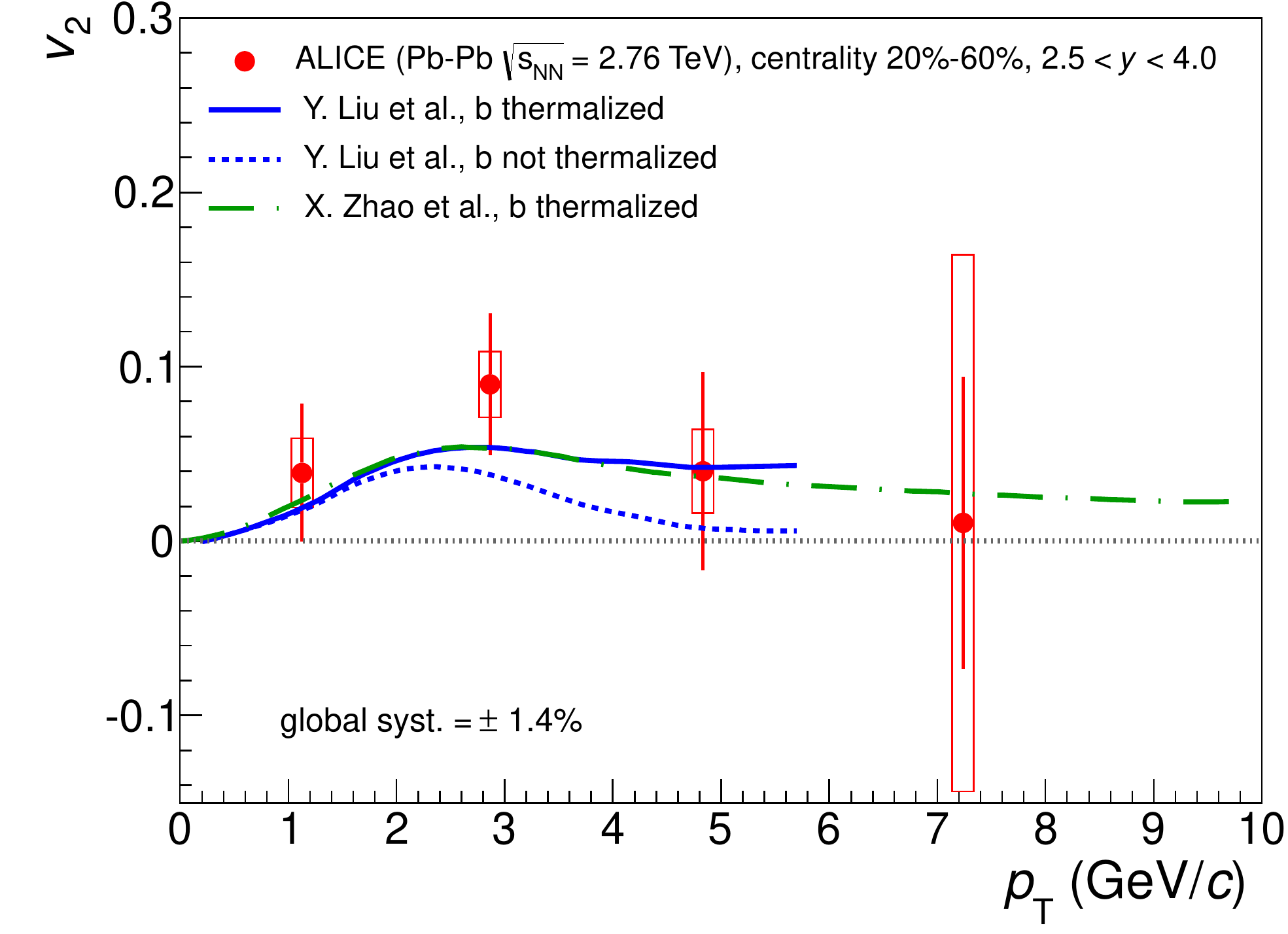}
\caption{Inclusive \jpsi $R_{\rm{AA}}$ as a function of the number of participants (left) and $v_2$ as a function of $p_T$ (right) at forward rapidity (from~\cite{jpsiflow}).}
\label{fig:jpsi}
\end{figure}

Nuclear modification factor for \jpsi mesons measured by ALICE at forward rapidity is shown in fig.~\ref{fig:jpsi} (left) as function of the number of participants $\langle  N_{\rm part}\rangle$~\cite{jpsiRaaForward}. The observed suppression flattens at $\langle  N_{\rm part}\rangle \sim 100$ and appears to be much weaker than at RHIC for central collisions. Such a centrality dependence and $p_{\rm T}$ differential studies suggest that $c\bar{c}$ regeneration processes indeed play an important role at LHC energies. The
observed hint for a non-zero elliptic flow shown in fig.~\ref{fig:jpsi} (right) is also in favour of this picture~\cite{jpsiflow}.

\begin{figure}[htb]
\centering
\includegraphics[width=10cm]{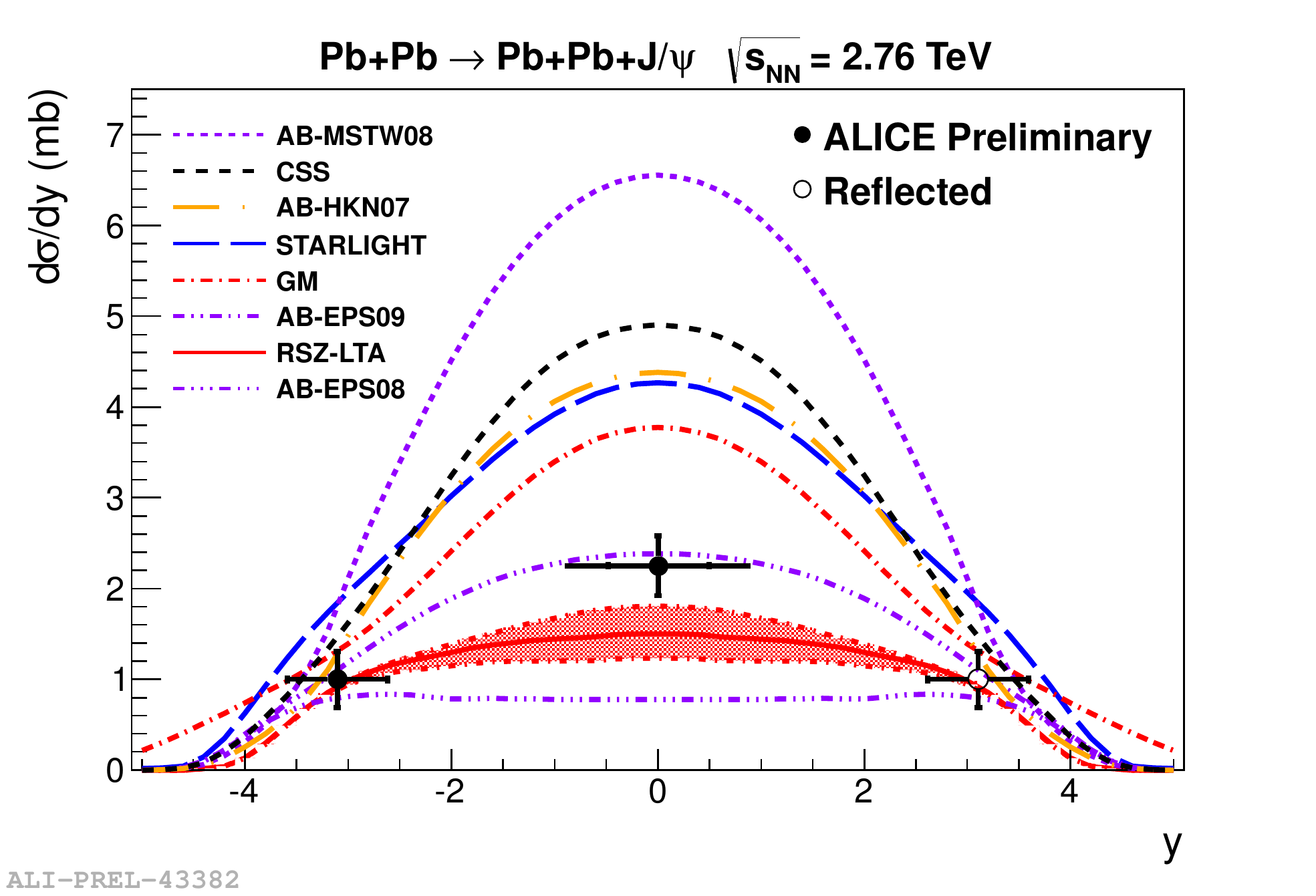}
\caption{Coherent \jpsi photoproduction cross section measured with ALCE and compared with model predictions.}
\label{fig:coherent}
\end{figure}

\jpsi production has been also studied in ultraperipheral collisions (UPC) characterized by impact parameters larger than two nuclear radii resulting in a strong suppression of hadronic processes and dominance of photon-induced reactions. Measurement of the coherent quarkonium photoproduction cross section in UPC is of particular interest since in the Leading Order pQCD it is proportional to the squared nuclear gluon density providing direct tool to study the nuclear gluon shadowing at small $x \sim 10^{-3}-10^{-2}$.  ALICE results on coherent \jpsi photoproduction at forward~\cite{coherent-forward} rapidity and preliminary results at central rapidity are compared with model predictions in fig~\ref{fig:coherent}. The best agreement is found for the model~\cite{ab} which incorporates nuclear gluon shadowing effects according to EPS09LO global fits~\cite{Eps09}.

\section{Proton-lead results}

Proton-nucleus collisions are very important to discriminate between initial state, cold nuclear matter effects in heavy ion collisions and effects of hot QCD matter dynamics. During a short pilot p--Pb run at~$\sqrt{s_{NN}} = 5.02$ TeV in September 2012, ALICE collected 0.9 $\mu$b$^{-1}$ of minimum bias triggers  and already published first  results on this data~\cite{pAmult,pArpA}.

\begin{figure}[b]
\includegraphics[width=7.15cm]{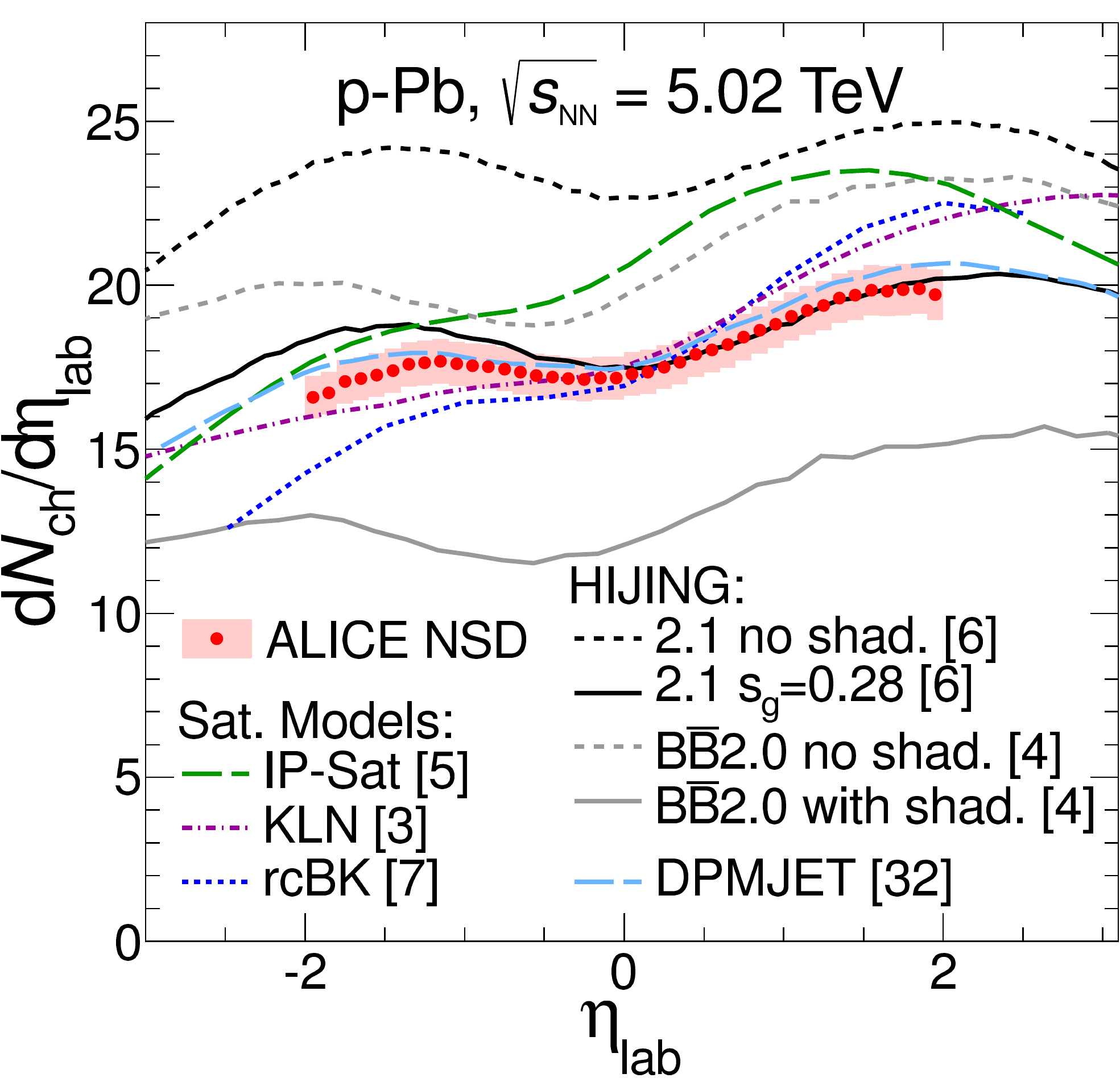}
\includegraphics[width=7.85cm]{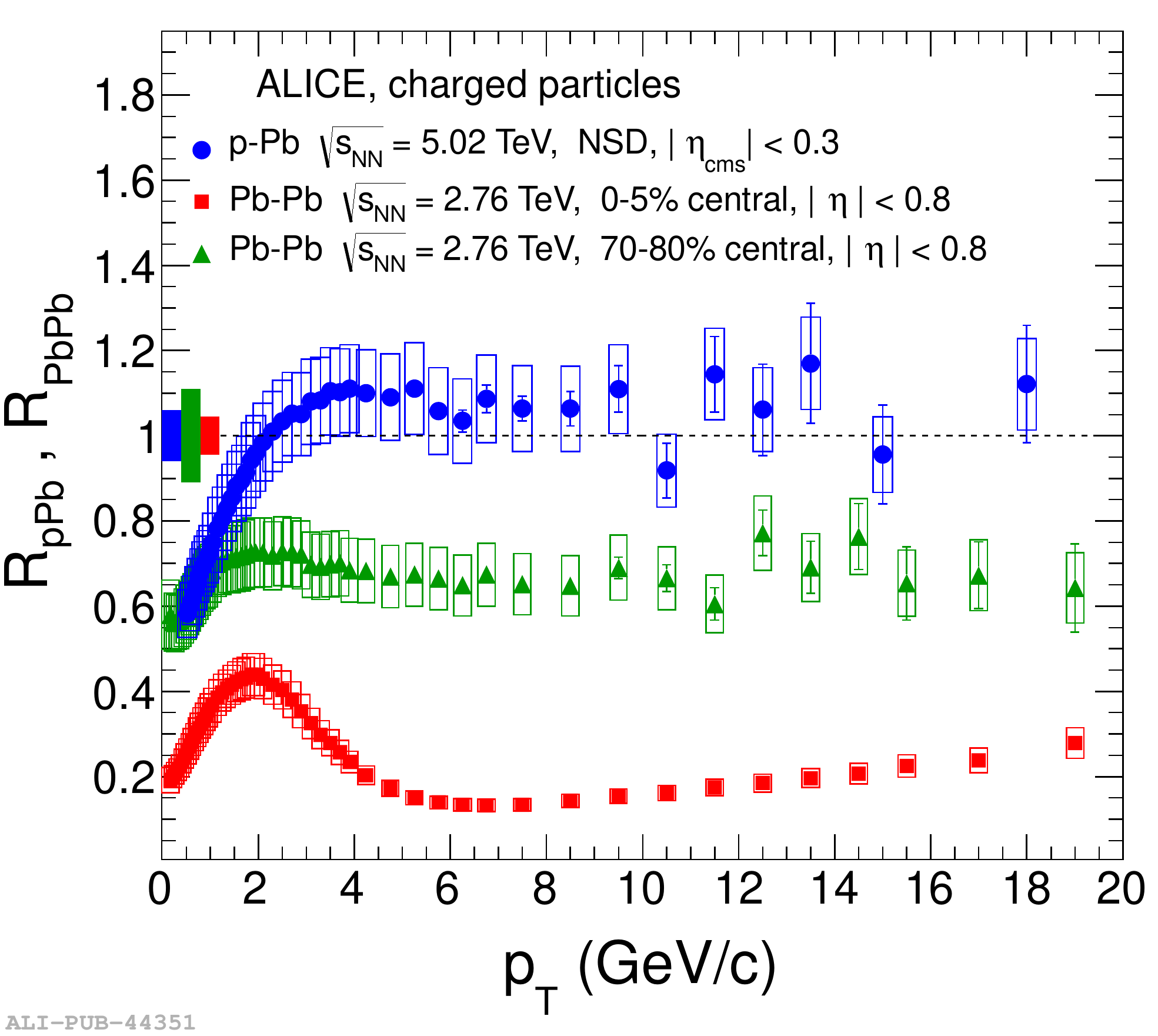}
\caption{Left: pseudorapidity density of charged particles measured
in non-single-diffractive p--Pb collisions at~$\sqrt{s_{NN}} = 5.02$ TeV compared to theoretical
predictions~\cite{pAmult}. Right: nuclear modification factor in p--Pb and Pb--Pb collisions at LHC~\cite{pArpA}.}
\label{fig:pA}
\end{figure}

Fig.~\ref{fig:pA} (left) shows the pseudorapidity dependence of the charged particle density in non-single-diffractive p--Pb events measured by ALICE~\cite{pAmult} in comparison with theoretical predictions. The best agreement is found for DPMJET model and HIJING2.1 with the shadowing parameter $s_g$ tuned to describe RHIC d--Au data while gluon saturation models predict a steeper pseudorapidity dependence. Midrapidity particle density was also scaled to the number of participants and compared to few measurements at lower energies showing the trend similar to pp energy dependence.

ALICE has also measured the nuclear modification factor $R_{\rm pPb}$ of charged particles as function of transverse momentum~\cite{pArpA}. Fig.~\ref{fig:pA} (right) presents the results in comparison to $R_{AA}$ in central and peripheral Pb--Pb collisions measured with ALICE at $\sqrt{s_{NN}} = 2.76$ TeV.  The observed $R_{\rm pPb}$ is consistent with unity at transverse momentum above 2 GeV indicating that the strong suppression of hadron production measured in Pb--Pb collisions at LHC is not an initial state effect but is a consequence of jet quenching in hot QCD matter. 

In the beginning of 2013, ALICE collected more than 30 nb$^{-1}$ of pA data allowing to make further progress in understanding of cold nuclear matter effects. 

\section{Upgrade plans}

LHC has recently entered a two-year period of the Long Shutdown 1 dedicated to consolidation to the full designed energy of 7 TeV per beam. ALICE will take this opportunity to complete the approved detector design and consolidate the detector performance. In 2015--2017, ALICE is going to fulfill initially approved plans and collect about 1 nb$^{-1}$ of Pb--Pb data at full LHC energy. This data taking period will be followed by Long Shutdown 2 aiming to upgrade LHC luminosity capabilities by factor 10 and allowing to reach Pb--Pb interaction rates of about 50 kHz. 

ALICE is going to take full advantage of the high-luminosity LHC. The intent is to explore $10^{10}$ minimum bias events, factor 100 more than planned before the upgrade, allowing to perform multi-dimensional analysis of rare low-$p_{\rm T}$ observables (e. g. charm baryons) with unprecedented statistical and systematic accuracy~\cite{upgrade-all}. This ambitious goal relies on the development of the new data acquisition system based on pipeline readout and the high-level trigger with full online event reconstruction and selection. In addition, readout electronics of all detector subsystems has to be upgraded and new GEM-based readout chambers have to be installed in TPC to meet the high rate requirements. Besides, in order to exploit low-$p_{\rm T}$ and charm physics, ALICE is going to install a smaller radius beam pipe and a new inner tracking system with pipeline readout, reduced material budget of only 0.3\% of radiation length and smaller pixel sizes providing hit resolution of the order of $4\times4 \mu$m and resulting in factor 3 improvement in the vertexing precision~\cite{upgrade-its}.

\section{Conclusion}
ALICE obtained a wealth of physics results from the first heavy ion runs at LHC. Now we are entering a charm era of precision measurements: we are looking forward to the new Pb--Pb data at higher energy with complete approved ALICE detector and have an ambitious upgrade program for the high luminosity LHC.

\end{document}